\documentstyle[preprint,aps]{revtex}
\tighten
\input epsf
\begin{document}
\draft
\preprint{
\begin{tabular}[t]{r}
 SI-97-07 \\
UCLA/97/TEP/9\\
\end{tabular}
}

\title{Dynamic SU(2) Lattice Gauge Theory\\
at Finite Temperature\thanks{
Work supported in part by the Deutsche Forschungsgemeinschaft; DFG~Schu
95/9-1}}

\author{K. Okano$^*$
\thanks{
On leave of absence from Tokuyama University,
 Tokuyama-shi, Yamaguchi 754, Japan}, 
 L. Sch\"ulke and B. Zheng}
\address{Universit\"at -- GH Siegen, D -- 57068 Siegen, Germany}
\address{$^*$ University of California Los Angeles, CA90095-1547, US}

\maketitle

\begin{abstract} 
The dynamic relaxation process for the 
(2+1)--dimensional
SU(2) lattice gauge theory at critical temperature
is investigated with Monte Carlo methods.
The critical initial increase of the Polyakov loop is 
observed.
 The dynamic exponents $\theta$ and $z$ as well as 
 the static critical exponent $\beta/\nu$ are determined
from the power law behaviour of the Polyakov loop,
the auto-correlation
and the second moment at the early stage of the time 
evolution.
The results are well consistent and
universal short-time scaling behaviour of the dynamic 
system
is confirmed.
The values of the exponents show that
the dynamic SU(2) lattice gauge theory is in the same 
dynamic universality
class as the dynamic Ising model.
\end{abstract}


In recent years lattice gauge theory has continuously been 
developing.
Numerical simulations have been extended to the gauge 
theory with fermions
and at finite temperature. 
Up to now, however, most of simulations 
are devoted to the equilibrium state.
Non-equilibrium processes have not been studied so deeply.
In this letter, we report the first numerical simulation
of the short-time dynamics for the
(2+1)-dimensional SU(2) lattice gauge theory
at the critical temperature. 

Recently great progress has been achieved
in critical dynamics for spin systems.
For long it was believed that no universal behaviour would 
be present
in the short-time regime of critical dynamics.
However, for the critical relaxation process starting from 
an initial state
with {\it very high temperature} and {\it small 
magnetization},
it was recently argued by Janssen, Schaub and Schmittmann 
\cite {jan89}
with renormalization group methods that
there exist
universality and scaling even {\it at macroscopic early 
times},
 which sets in right after a
microscopic time scale $t_{mic}$.
Based on the scaling relation it was
predicted that at the beginning of the time evolution the 
magnetization
 surprisingly
undergoes a {\it critical initial increase}
\begin{equation}
M(t) \sim m_0 \, t^\theta
\label{e10}
\end{equation}
where $\theta$ is a new dynamic exponent.

Numerical simulations 
 support the above predictions.
The critical initial increase of the magnetization in 
Eq.~(\ref{e10}) was observed
for the Ising model and the Potts model and the exponent 
$\theta$
was directly measured \cite {li94,sch95}.
The scaling relation
and universality are confirmed \cite 
{hus89,hum91,li94,men94,oka97a,oka97}.
The microscopic time scale $t_{mic}$ is around
$5-60$ Monte Carlo time steps, depending on the 
observables and the 
microscopic details. If one argues that a Monte Carlo time
step is a typical microscopic time unit, such a result
is reasonable.
Compared with in the long-time regime,
the scaling variable $t$ in short-time dynamics 
plays a more important role in understanding the scaling
behaviour, especially in numerical simulations.
The investigation of the universal behaviour of
the short-time dynamics not only enlarges the fundamental 
knowledge
on critical phenomena but also, more interestingly,
provides possible new ways to determine all the
dynamic exponents as well as the static exponents
from the short-time dynamics,
either based on the power law behaviour of the observables 
at the beginning
of the time evolution \cite {sch95,gra95,oka97a,cze96},
or on finite size scaling \cite {li95,li96}. 
An appealing feature of these methods is that they
may be free of critical slowing down

At this stage, it is natural to ask whether such an 
investigation may
be generalized to field theory. This would be very 
important
for the understanding of {\it the non-equilibrium 
properties
of field theory and numerical simulations of lattice gauge 
theory}.
On the other hand, the two--dimensional Ising and 3-state 
Potts model are
known to be the simplest models presenting critical 
phenomena.
They are known to  have quite clean behaviour
in many respects. One may wonder whether nice universal
short-time behaviour is special for these simple systems.
For example, it could be that the microscopic time scale 
$t_{mic}$
for more complicated systems is so big such that
it is comparable with the macroscopic time scale.
Then we can not observe
any universal behaviour in the macroscopic short-time 
regime.

As a first approach to the {\it dynamic } gauge theory,
 we will numerically
investigate the (2+1)--dimensional SU(2) lattice
gauge theory at finite temperature with a dynamics of model A
\cite {hoh77} even though there exists so far not
any analytical investigation. The motivation to choose this 
model is that the deconfining phase 
transition observed in this model is the second order.
 On the other hand, it is expected that in 
equilibrium
 the (2+1)--dimensional SU(2) lattice gauge theory is in the same 
universality class as the two--dimensional
Ising model \cite {sve82}.
 This has also been numerically observed \cite{chr91,tep93}.
It is interesting to investigate whether the dynamic SU(2) 
lattice gauge theory
is also in the {\it same dynamic universality class} as 
the dynamic Ising model.

The (2+1)--dimensional SU(2) lattice gauge theory 
is described by the Hamiltonian
\begin{equation}
H= - \frac {4}{g^2}  \sum_{P} U_P
\label{e20}
\end{equation}
where the index $P$ indicates the sum over all the 
fundamental plaquettes
\begin{equation}
U_P=U_{\mu}(x)U_{\nu}(x+\hat \mu)U^{\dagger}_{\mu}(x+\hat 
\mu+\hat \nu)
   U^{\dagger}_{\nu}(x+\hat \nu)
\label{e30}
\end{equation}
with $\mu$, $\nu$ denoting the directions in the ($2$+$1$) 
dimensional space.
As {\it a finite temperature theory},
the lattice size should be taken to be 
$N^2$$\times$$N_{0}$.
Here $N_{0}$ corresponds to the inverse temperature.
Since the theory is super-renormalizable, the continuum
scaling law is of the simple form $N_{0} T \sim g^{-2}$.
For this SU(2) lattice gauge theory 
in equilibrium there exist
already rather good numerical results.
For a lattice with $N=64$ and $N_{0}=2$,
Christensen and Damgaard obtained
the critical point $4/g^2_c=3.39$ and the exponent 
$\beta=0.120(8)$ \cite {chr91}, while Teper got the 
critical point 
$4/g^2_c=3.47$ and the exponent $\nu=0.98(4)$ \cite 
{tep93}.
Compared with the exact values $\beta=0.125$ and $\nu=1.0$
for the two--dimensional Ising model, the numerical results
of the critical exponents
for the SU(2) lattice gauge theory support that
both models are in the same universality class.
Values for the critical point from the two groups of authors
show some small difference. The determination
of the critical point to a very rigorous level is at the 
present
stage still difficult. For simplicity, in this paper we 
will
take the simple average $4/g^2_c=3.43$ of
the above given two values
as an input. Our numerical results
show that this value for the critical point is very 
close to the real one.

In order to simulate a critical relaxation,
one should first prepare an initial state.
The magnetization in the SU(2) lattice gauge theory
is defined as the globally averaged
Polyakov loop
\begin{equation}
M (t) = \frac {1}{N^2}  \sum_{i} < W_i (t) >,
\label{e40}
\end{equation}
where $W_i$ is the Polyakov loop at site $i$ which
locates in the two--dimensional lattice 
with lattice size $N^2$. 
The average $<...>$ is over the random forces and
the independent initial configurations.
Compared with the Ising model
 the Polyakov loop $W_i$ plays the role of the Ising
spin $S_i$ 
\footnote {However, $W_i$ in the SU(2) theory
is a real variable defined
in the interval $[-1,1]$ while the spin $S_i$ takes 
integers $\pm 1$.
Therefore the SU(2) lattice gauge theory should contain
more physical contents. But in this paper we  will not
discuss this in details.}.
Following the idea of Janssen, Schaub and Schmittmann 
\cite {jan89},
the initial state should have zero
spatial correlation length and 
small initial magnetization.
To get initial configurations with zero spatial correlation,
we just remember that for fixed $N_0$,
the center symmetry $Z_2$ of the SU(2) theory is broken for 
large $4/g^2$,
while it remains unbroken for small $4/g^2$.
Therefore, for the initial state we take the coupling
 $4/g^2=0$. A non-zero initial magnetization can be 
achieved in many ways. A natural one is to introduce 
an initial external
magnetic field $h$. Summarizing, the initial configurations
 can be generated
by the initial Hamiltonian $H_0=h \sum_{i} W_i$.
Final procedure is to adjust the
configuration so generated as to give the initial
magnetization sharply. Different methods can be found for
this sharp preparation in literature in statistical physics
\cite {li94,sch95,oka97a,oka97,maj96a,sch97}.

After an initial configuration is prepared, the system is 
suddenly
quenched to the critical temperature with the Hamiltonian
in Eq. (\ref {e20}) and then released to evolve
with a dynamics of model A. In this paper we adopt the
{\it heat-bath algorithm} for the dynamic evolution.
We stop the update of the dynamic system at a reasonable 
time,
which is typically some hundreds Monte Carlo time steps,
and repeat the process. The average is taken over 
different 
independent initial configurations and random numbers.
The total sample for the lattice size up to $N=64$ 
is $9\ 600$ while for $N=128$ is $8\ 000$.
Errors are estimated by dividing the total sample into 
three
 or five groups.

\begin{figure}[p]\centering
\epsfysize=12cm
\epsfclipoff
\fboxsep=0pt
\setlength{\unitlength}{1cm}
\begin{picture}(13.6,12)(0,0)
\put(1.9,2.0){\makebox(0,0){0.01}}
\put(1.9,11.0){\makebox(0,0){0.1}}
\put(1.2,8.0){\makebox(0,0){$M(t)$}}
\put(9.8,1.2){\makebox(0,0){$t$}}
\put(7.,7.9){\makebox(0,0){\footnotesize$m_0=0.04$}}
\put(12.,7.8){\makebox(0,0){\footnotesize$N=16$}}
\put(4.5,7.3){\makebox(0,0){\footnotesize solid line: $N=64$}}
\put(8.,3.5){\makebox(0,0){\footnotesize$m_0=0.02$,\quad$N=64$}}
\put(11.5,9.4){\makebox(0,0){\footnotesize$N=32$}}
\put(0,0){{\epsffile{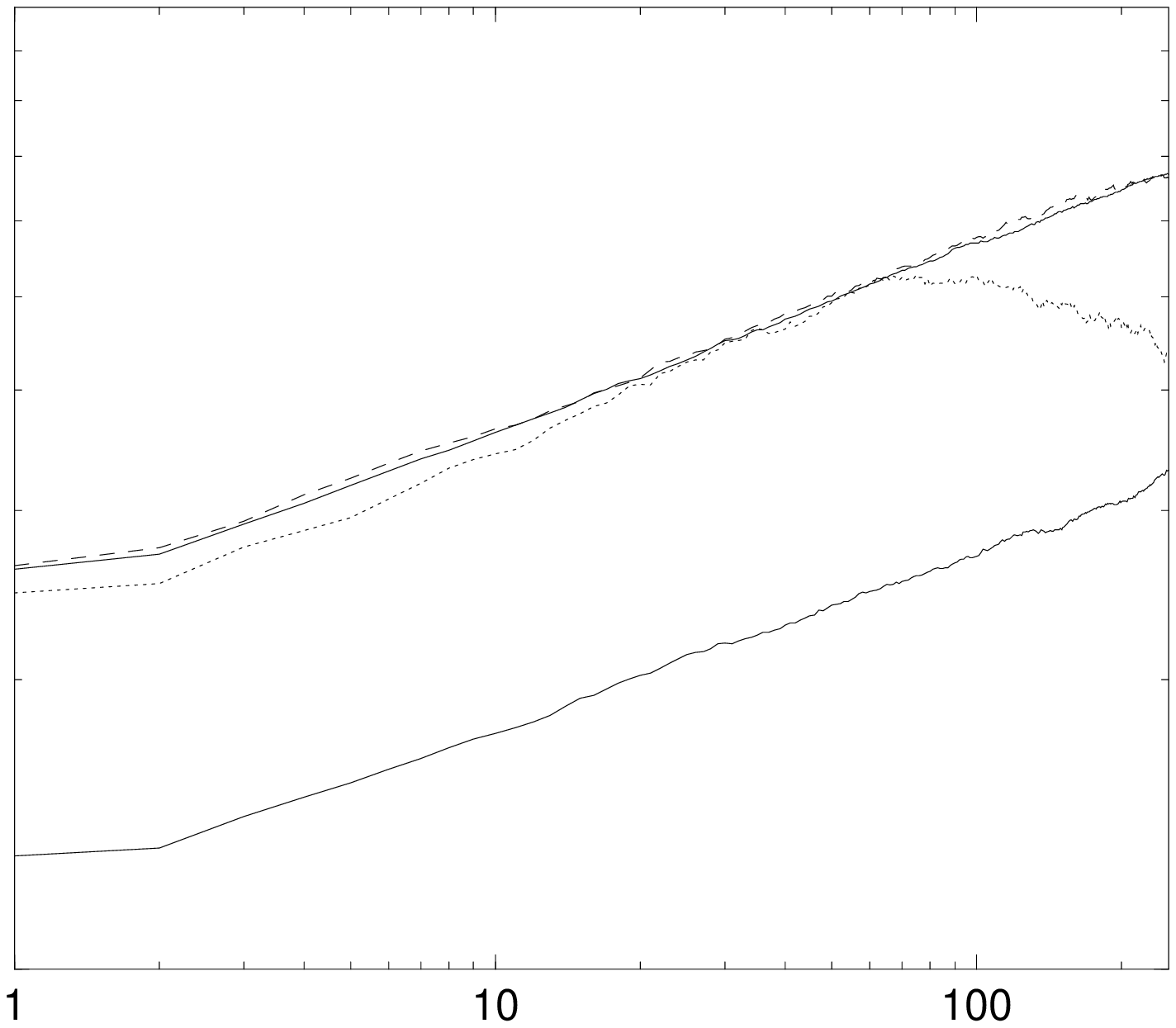}}}
\end{picture}
\caption{ The time evolution of the magnetization
for different lattice sizes $N$ and initial magnetization 
$m_0$
with the heat-bath algorithm is plotted in double-log 
scale.
}
\label{f1}
\end{figure}

In Fig.~\ref {f1}, time evolution of the magnetization is 
plotted
in double-log scale
for different initial magnetization $m_0$ and lattice 
sizes $N$.
The solid line above is the magnetization profile for 
$m_0=0.04$ and $N=64$, while the dotted and dashed line
are those for $N=16$ and $32$ respectively.
The solid line below is the time dependent magnetization
with $m_0=0.02$ and $N=64$. From the figure one can 
realize
that in the {\it first} Monte Carlo time step
the magnetization drops. For example, for 
$M(t=0)=m_0=0.04$ 
with lattice size $N=64$ it drops to $M(t=1)=0.026$.
Such a dropping within the microscopic time scale 
$t_{mic}$
is a typical non-universal behaviour which essentially 
depends
on the microscopic details. Similar phenomena have also 
been
observed for the two--dimensional Potts model and the XY model
\cite {oka97a,oka97}.
With the Metropolis algorithm for the Potts model,
the magnetization decreases
continuously even up to around $10$ Monte Carlo time 
steps.
From Fig.~\ref {f1}, we clearly see that after one Monte 
Carlo
time step, the magnetization indeed increases and for big 
enough
lattice sizes it is quickly
stabilized to the universal power law behaviour given 
in Eq. (\ref {e10}).
The microscopic time scale is $t_{mic} \sim 20$.
For $m_0=0.04$, the magnetization profiles
for $N=32$ and $64$ do not show a big difference. 
The finite size effect for $N=64$ is already quite small. 
From the slope of the curves in Fig.~\ref {f1},
one can measure the critical exponent $\theta$.
For $m_0=0.04$ and $N=64$, from a time interval
$[20,250]$ we obtain the exponent $\theta = 0.192(2)$.
This value is very well consistent with that for the two 
dimensional
Ising model \cite {oka97a,gra95}.
 The comparison can be done in Table~\ref {t1}.
In Table~\ref {t1}, the value of the exponent $\beta/\nu$ 
for the Ising model
is exact 
while those of the exponents $\theta$ and $z$ are taken 
from the literature \cite {oka97a}.
Rigorously speaking, the critical exponent $\theta$ 
is defined in the limit $m_0=0$. A numerical measurement
in this limit is practically not possible.
Therefore, a linear extrapolation to the limit $m_0=0$ 
from
finite $m_0$ should in principle 
be carried out \cite {sch95,oka97a}. 
For this reason, we have also performed the simulation
for $m_0=0.02$ and $N=64$. Since $m_0$ now is smaller,
the fluctuations become bigger. The measured exponent is
$\theta =0.186(12)$. Within the errors 
we can not distinguish the results for 
 $m_0=0.02$ and $m_0=0.04$. Therefore in this paper a
linear extrapolation will not be performed.

Now we set $m_0=0$ and proceed to measure the 
auto-correlation
$A(t)$ and the second moment $M^{(2)}(t)$. 
From a careful scaling analysis for Ising-like systems 
\cite {jan92,jan89}, we expect the short-time dynamic 
scaling behaviours
\begin {equation}
A(t) \equiv \frac {1}{N^2} \sum_{i} < W_i (t) W_i(0) >
         \sim t^{\theta - d/z}
\label {50}
\end {equation}

\begin {equation}
M^{(2)}(t) \equiv \frac {1}{N^4} < ( \sum_{i} W_i (t) )^2 >
           \sim t^{(d-2\beta/\nu)/z}
\label {60}
\end {equation}
Important is here that the exponent $\theta$ also enters
the auto-correlation $A(t)$. Actually the first numerical
estimate of the exponent $\theta$ for the Ising model is
from the measurement of $A(t)$ by taking $z$ as an input
\cite {hus89,hum91}.
In this case, however, the error induced by $z$
is relatively  big since
the dynamic exponent $z$ is usually not known so 
rigorously
and the ratio $d/z$ is much bigger than $\theta$.
In contrast to this, with the exponent $\theta$ obtained
from the initial increase of the magnetization,
one can get a rather accurate value for the exponent $z$
from the auto-correlation. Compared with the traditional
measurement of $z$ from the exponential decay of the 
auto-correlation
in the long-time regime, to some extent, our short-time
dynamic approach
is free of critical slowing down. Due to the large time 
correlation
length, for big lattice sizes the traditional measurement
is very difficult since independent configurations
can hardly be generated. However, in our short-time 
dynamic approach
the measurement is always carried out in the short-time 
regime,
where the system rather rapidly converges to the universal
power law behaviour as the lattice size increases.
In principle, we do not have the problem how to generate
independent configurations efficiently.
On the other hand, in the dynamic approach the average
is really the sample average rather than the time average
based on the ergodicity assumption. This fact may show
its merit in future.

Here we should stress that the dynamic exponent $z$
is originally defined in the long-time regime of the 
dynamic
evolution. It is a conceptual progress that
we can determine it from the universal short-time
behaviour. More importantly, not only the dynamic exponent
$z$ but actually also the static exponents, which are
defined in equilibrium state, can numerically be measured
from the short-time dynamics, e.g. from the second moment
$M^{(2)}(t)$.

\begin{figure}[p]\centering
\epsfysize=12cm
\epsfclipoff
\fboxsep=0pt
\setlength{\unitlength}{1cm}
\begin{picture}(13.6,12)(0,0)
\put(1.7,11.0){\makebox(0,0){0.01}}
\put(1.7,5.8){\makebox(0,0){0.001}}
\put(11.8,1.2){\makebox(0,0){$t$}}
\put(6.5,9.){\makebox(0,0){\footnotesize$A(t)$}}
\put(7.5,4.5){\makebox(0,0){\footnotesize$M^{(2)}(t)$}}
\put(0,0){{\epsffile{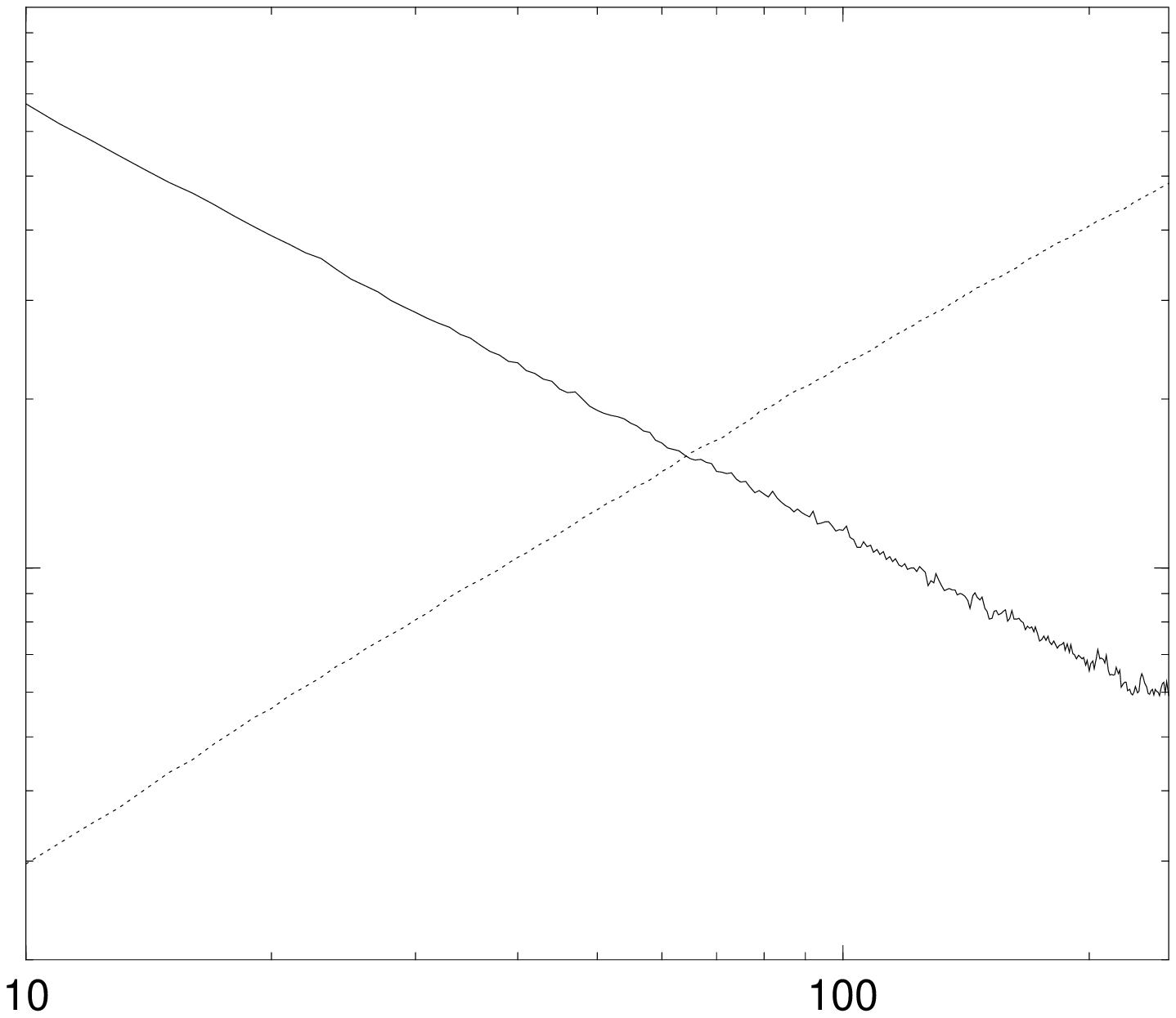}}}
\end{picture}
\caption{ The auto-correlation $A(t)$ and the second 
moment
$M^{(2)}(t)$ for the lattice sizes $N=128$ and
 initial magnetization $m_0=0$ obtained
with the heat-bath algorithm is plotted in double-log 
scale.
}
\label{f2}
\end{figure}

To measure the auto-correlation $A(t)$ and the second
moment $M^{(2)}(t)$, we have performed the simulation of $m_0=0$
with a lattice size $N=128$. In Fig.~\ref {f2}, 
$A(t)$ and $M^{(2)}(t)$
 are plotted in double-log scale with the solid
line and dotted line respectively.
After a microscopic time scale $t_{mic} \sim 50$,
nice power law behaviour is observed. In the numerical 
simulations
for the Ising model and Potts model, it is known that 
the microscopic time scale for the second moment is 
somehow
longer than that for the magnetization \cite {oka97a}.
This is also the case for the SU(2) lattice gauge theory.
We have performed a power law fit in the time interval 
$[60,250]$
and obtain the exponent $\theta - d/z=-0.745(12)$ while
$(d-2\beta/\nu)/z=0.825(14)$. These results are consistent 
with 
those of the two--dimensional Ising model,
$\theta - d/z=-0.737(01)$ and 
$(d-2\beta/\nu)/z=0.817(07)$.
Taking the exponent $\theta$ as an input, we can calculate
the values for the dynamic exponent $z$ and the static
exponent $\beta/\nu$. All these results and, for comparison
those of the Ising model are given in Table~\ref {t1}.
Even though the dynamic exponent $z$ has been known
for a long time, for the lattice gauge theory our 
measurement is
the first reliable one. The measured value 
$\beta/\nu=0.120(18)$
is also in agreement with $\beta/\nu=0.120(08)$  obtained 
in reference \cite {chr91},
and with the exact value  $\beta/\nu=0.125$
for the two--dimensional Ising model.

In conclusion, we have investigated the universal 
short-time behaviour
of the (2+1)--dimensional dynamic SU(2) lattice gauge theory. 
The critical initial increase of the Polyakov loop is 
observed.
The dynamic exponents $\theta$ and $z$ as well as
 the static critical exponent $\beta/\nu$ are determined 
from the power law behaviour of different observables at 
the
early stage of the time evolution. The values are 
consistent 
within the errors with those of the two-dimensional Ising model.
The results strongly support that there exists an 
universal
short-time scaling behaviour for the dynamic SU(2) lattice
gauge theory, and also suggest that 
  the (2+1)-dimensional dynamic SU(2) lattice gauge theory and 
the two-dimensional
Ising model
are in the same dynamic universality class.

{\it Acknowledgement:}
One of the authors (K.O) would like to acknowledge the
hospitality of the Theory Group of The Elementary Particle
Physics, University of California, Los Angeles for his
stay for one year during when the paper has been
completed.

%\bibliographystyle{stybase/pr_np}
%\bibliography{stybase/ising}

%\bibliographystyle{prsty}
%\bibliography{/ising} 

\begin{table}[p]\centering
\begin{tabular}{|c|l|l|l|}
    & $\theta$ & z & $\beta/\nu$ \\
\hline
 SU(2) & 0.192(02) & 2.135(27) & 0.120(18) \\
\hline
 Ising & 0.191(01) & 2.155(03) & 0.125 \\
\end{tabular}
\caption{ The exponents $\theta$, $z$ and $2\beta/\nu$ 
measured from the short-time dynamics with the heat-bath algorithm.}
\label{t1}
\end{table}

\end{document}